\documentclass[manuscript]{aastex}

\usepackage{rotating}

\newcommand{\kms}{km\,s$^{-1}$}

\newcommand{\visir}{\emph{VISIR}}

\newcommand{\flux}{ergs\,s$^{-1}$\,cm$^{-2}$}

\newcommand{\Av}{A$_{\rm v}$}

\newcommand{\myemail}{claire.martin-zaidi@obs.ujf-grenoble.fr}
\shorttitle{H$_2$ emission in the disk of HD~97048}
\shortauthors{Martin-Za{\"\i}di et al.}

\begin{document}


\title{Molecular hydrogen in the disk of the Herbig Ae star \object{HD~97048}}


\author{C. Martin-Za{\"\i}di\altaffilmark{1},
  E. Habart\altaffilmark{2}, J.-C. Augereau\altaffilmark{1},
  F. M\'enard\altaffilmark{1}, P-.O. Lagage\altaffilmark{3},
  E. Pantin\altaffilmark{3} and J. Olofsson\altaffilmark{1}}
%\email{claire.martin-zaidi@obs.ujf-grenoble.fr}

\altaffiltext{1}{Laboratoire d'Astrophysique de Grenoble, CNRS/UJF -
  UMR 5571, 414 rue de la Piscine, DU Saint Martin d'H\`eres, 38041
  Grenoble cedex 9, France, \myemail, augereau@obs.ujf-grenoble.fr,
  menard@obs.ujf-grenoble.fr, olofsson@obs.ujf-grenoble.fr}

\altaffiltext{2}{Institut d'Astrophysique Spatiale, 91405 Orsay,
  France, emilie.habart@ias.u-psud.fr}

\altaffiltext{3}{Laboratoire AIM, CEA/DSM - CNRS - Universit\'e Paris
  Diderot, DAPNIA/Service d'Astrophysique, Bat. 709, CEA/Saclay, 91191
  Gif-sur-Yvette Cedex, France, pierre-olivier.lagage@cea.fr,
  epantin@cea.fr}


\begin{abstract}
  We present high-resolution spectroscopic mid-infrared observations
  of the circumstellar disk around the Herbig Ae star HD~97048
  obtained with the {\it VLT Imager and Spectrometer for the
    mid-InfraRed (VISIR)}. We conducted observations of mid-infrared
  pure rotational lines of molecular hydrogen (H$_2$) as a tracer of
  warm gas in the disk surface layers. In a previous paper, we
  reported the detection of the S(1) pure rotational line of H$_2$ at
  17.035 $\mu$m and argued it is arising from the inner regions of the
  disk around the star. We used \visir\ on the VLT for a more
  comprehensive study based on complementary observations of the other
  mid-infrared molecular transitions, namely S(2) and S(4) at 12.278
  $\mu$m and 8.025 $\mu$m respectively, to investigate the physical
  properties of the molecular gas in the circumstellar disk around
  HD~97048. We do not detect neither the S(2) line nor the S(4) H$_2$
  line from the disk of HD~97048, but we derive upper limits on the
  integrated line fluxes which allows us to estimate an upper limit on
  the gas excitation temperature, T$_{ex}<570$~K. This limit on the
  temperature is consistent with the assumptions previously used in
  the analysis of the S(1) line, and allows us to set stronger
  contraints on the mass of warm gas in the inner regions of the
  disk. Indeed, we estimate the mass of warm gas to be lower than 0.1
  M$_{\rm Jup}$. We also discuss the probable physical mechanisms
  which could be responsible of the excitation of H$_2$ in the disk of
  HD~97048.
\end{abstract}


\keywords{stars: pre-main sequence --- stars: individual (HD97048) ---
  (stars:) circumstellar matter --- (stars:) planetary systems:
  protoplanetary disks --- infrared: stars}



\section{Introduction}

Disks around young stars are a natural outcome of the star formation
process and the place for planet formation. At the present time, a
significant effort has been done on the study of the dust in
disks. However, the dust only represents a tiny fraction of the disk
mass ($\sim$1\%), and it is thus mandatory to deeply study the gas
phase in disks in order to set stronger constraints on the giant
planets formation process. Molecular hydrogen (H$_2$) is the most
abundant molecule in the circumstellar (CS) environments of young
stars and is supposed to be the key element of giant planet formation,
thus its diagnostics are promising. Indeed, the detection of H$_2$
provides the most direct information about the gaseous content of
disks, setting limits on the timescales for the dissipation of CS
matter and possibly planet building. H$_2$ has been observed in CS
environments at ultraviolet \citep[e.g.][]{Johns-Krull00, klr08a} and
near-infrared \citep[e.g.][]{Bary03} wavelengths. These observations
trace hot circumstellar gas, or gas excited by fluorescent processes,
or require specific spatial distributions for the gas to be
detectable. They are therefore difficult to translate into gas
masses. The pure rotational mid-infrared H$_2$ lines are useful probes
because the level populations are expected to be in local
thermodynamic equilibrium (LTE) at the local gas temperature, and so
line ratios allow determination of the excitation temperature and mass
of the warm gas.

\object{HD~97048} is a nearby, relatively isolated Herbig A0/B9 star,
surrounded by a CS disk, located in the Chameleon cloud at a distance
of 180 pc \citep{VdA98}. Its age has been estimated from evolutionary
tracks to be of the order of 3 Myrs (kindly computed for us by
L. Testi and A. Palacios). The \visir\ \citep[{\it VLT Imager and
  Spectrometer for the mid-InfraRed};][]{Lagage04} imaging
observations of this star have revealed emission of PAHs (Polycyclic
Aromatic Hydrocarbons) at the surface of a flared disk extending at
least up to 370 AU \citep{Lagage06}. The flaring index has been
measured to be 1.26$\pm$0.05, which is in good agreement with
hydrostatic flared disk models \citep{Lagage06, Doucet07}.  This is
the only Herbig star for which the flaring of the disk has been
observed by direct imaging. This geometry implies that a large amount
of gas should be present to support the hydrostatic structure and that
the disk is at an early stage of evolution.  This star is thus one of
the best candidates to study the gas component in the disks of HAes.

In a previous paper \citep[][hereafter Paper~I]{klr07a}, we presented
\visir\ high-resolution spectroscopic mid-infrared observations of the
circumstellar disk around the Herbig Ae (HAe) star HD~97048 near 17.03
$\mu$m. Although \cite{Carmona08} suggested, from their disk model
with the assumptions of LTE conditions and gas-to dust ratio of about
100, that mid-IR H$_2$ lines should not be detected with the existing
instruments, we detected the S(1) pure rotational line of molecular
hydrogen at 17.035 $\mu$m arising from the disk around HD~97048. In
addition, \cite{Bitner07} have detected mid-IR H$_2$ rotational lines
in the disk of another Herbig star, namely AB Aur, using the high
spectral and spatial resolution TEXES spectrometer. These detections
demonstrate that H$_2$ can be observed in the mid-IR domain when
particular physical conditions exist in disks.

Despite the fact that the line is neither spatially nor spectrally
resolved, the detection of the S(1) line in the disk of HD~97048
revealed the presence of significant amounts of warm gas in the inner
35 AU of the disk. Circumstellar gas had been previously detected in
the disk of HD~97048 by \cite{Acke06}, who showed that the [O~I]
emission arises at radii between 0.5 and 60 AU from the central
star. Very recently, modeling of near-infrared CO emission showed that
the inner radius of the CO emitting region is located at 12 AU from
the star \citep{VdPlas_08b}. These observations reinforced the claim
that the disk of HD~97048 contains large amounts of gas both at small
and large radii, as suggested by the flaring geometry, and that
HD~97048 is a young object surrounded by a disk at an early stage of
evolution.  Indeed photoevaporation of the gas and planet formation
clear up the inner part of the disk in timescales expected to be lower
than 3 million years \citep{Takeuchi05, Gorti08a, Gorti08b}. The H$_2$
S(1) line detection also implies that particular physical conditions
for H$_2$ are present in the inner disk surface layer. The analysis of
these data usually assumes that the H$_2$ excitation is in local
thermodynamic equilibrium (LTE), and can itself thus be characterized
by a single excitation temperature, which should be close to the gas
temperature because of the low critical densities. As shown by the AB
Aur observations \citep{Bitner07, Bitner08}, the H$_2$ gas temperature
could be significantly higher than the dust temperature in the disk
surface layers of HD~97048, due to the dust settling or coagulation,
for example. However, several competing mechanisms could contribute to
the excitation of molecular hydrogen such as UV and X-rays heating,
shocks, etc..., \citep[see review papers by][]{Habart04b, Snow06} and
could be responsible for the observed emission. We refer the reader to
Paper~I for a complete discussion about the different possible
explanations.

The detection of the other pure rotational lines of H$_2$ is a
potentially powerful tool to determine the excitation temperature (and
thus the mass) of the warm gas, and better constrain the excitation of
H$_2$. In this paper, we thus present \visir\ high-resolution
spectroscopic observations of the two other pure rotational lines of
H$_2$ observable from the ground: the S(2) line at 12.2786 $\mu$m, and
the S(4) line at 8.0250 $\mu$m. We also recall the main results
obtained from the S(1) line detection (detailed in Paper~I).

%________________________________________________________________
\section{Observations and data reduction}
\label{reduction}

HD~97048 was observed at 3 different epochs. The observations at
17.035 $\mu$m presented in the Paper~I were performed in 2006 June 22,
the 8.025 $\mu$m observations in 2007 April 07, and the 12.278 $\mu$m
observations in 2007 July 03. The three sets of observations were
obtained with the high-resolution spectroscopic mode of \visir. The
exposure time, slit width, and atmospheric conditions during the
observations for the target star and standard references are presented
in Table~\ref{tab_obs}.

For all the three observations, the standard chopping and nodding
technique was used to suppress the large sky and telescope background
dominating at mid-IR wavelengths. Asteroids and standard stars were
observed just before and after observing HD~97048, at nearly the same
airmass and seeing conditions as the object. In order to correct the
spectra from the Earth's atmospheric absorption, we divided each
spectrum of HD~97048 by that of the corresponding asteroid (which has
a much better signal-to-noise ratio than that of the standard star),
and used the standard stars observed and modeled spectra
\citep{Cohen99} to obtain the absolute flux calibration. In order to
verify our flux calibration, we checked the {\it Spitzer Space
  Telescope} archives. HD~97048 was observed with the IRS spectrograph
installed onboard the {\it Spitzer Space Telescope} as part of the
``Cores to Discs'' (c2d) legacy program (AOR:0005644800, PI:
Evans). The data reduction was performed using the {\it c2d} legacy
team pipeline \citep{Lahuis06} with the S15.3.0 pre-reduced (BCD)
data. The absolute flux calibration of our \visir\ data in the 3
wavelength ranges has been performes in order to be consistent with
the {\it Spitzer} flux measurements. We estimate the error on the
absolute flux calibration to be lower than 10\% , a value which does
not affect our analysis.

The wavelength calibration is done by fitting the observed sky
background features with a model of Paranal's atmospheric emission
(for more details on the observation and data reduction techniques see
Paper~I). We note that \Av = 0.24 mag for HD~97048 \citep{Valenti00},
thus we have not corrected the spectra for dust extinction, since it
is negligible in our wavelength range for any \Av $<$ 40 mag
\citep{Fluks94}.

%________________________________________________________________
\section{Results}
\label{analysis}

\subsection{Non-detection of the S(2) and S(4) lines}

\begin{figure*}[!htbp]
\begin{center}
\includegraphics[width=16cm]{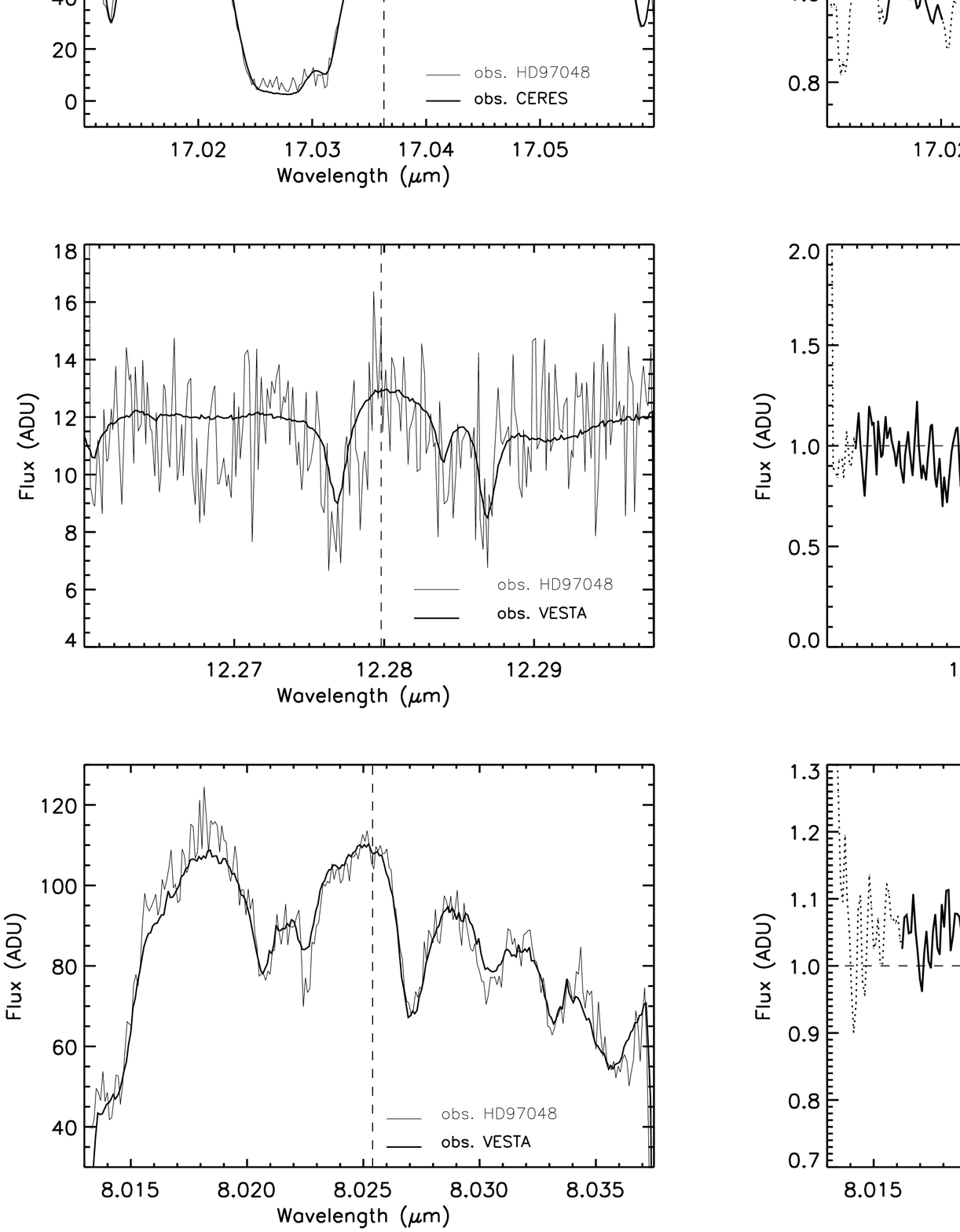}
\caption {\visir\ spectra of HD~97048 at 17.035 $\mu$m ({\it top
    panel}), 12.278 $\mu$m ({\it middle panel}) and 8.025 $\mu$m ({\it
    bottom panel}).  {\it Left panel:} continuum spectra of the
  asteroid and of the target before telluric correction. {\it Central
    panel:} full corrected spectra: dotted lines show spectral regions
  strongly affected by telluric features. {\it Right panel:} zoom of
  the region where the H$_2$ lines should be observed (dashed vertical
  lines).  The spectra were corrected neither for the radial velocity
  of the targets nor the Earth's rotation velocity.}
\label{HD97}
\end{center}
\end{figure*}

HD~97048 spectra show no evidence for H$_2$ emission neither at 12.278
$\mu$m nor at 8.025 $\mu$m (Fig.~\ref{HD97}, middle and bottom
panels). For each flux-calibrated spectrum, we calculated the standard
deviation ($\sigma$) for wavelength ranges relatively unaffected by
telluric absorption, and close to the wavelength of interest. The
3$\sigma$ upper limits on the integrated line fluxes were calculated
by integrating over a Gaussian of FWHM equal to a spectral resolution
element (21 \kms) and an amplitude of about 3$\sigma$ flux
(Fig.~\ref{HD97}). We assumed the same radial velocity for the S(2)
and S(4) lines than that observed for the S(1) line and corrected
appropriately for each epoch of observation. We thus centered the
Gaussian on the expected wavelengths for the S(2) and S(4) lines
respectively (see Table~\ref{tab_flux}). From the limits on integrated
intensities, we estimated the upper limits on the column densities of
the corresponding upper rotational levels of each H$_2$ transition
(see Table~\ref{tab_flux}). For this purpose, we first assumed that
the line is optically thin and that the radiation is isotropic. In
this context, the column densities are derived from the following
formula \citep{Van_Dishoeck_92}:

\begin{eqnarray}
I_{ul}=\frac{hc}{4 \pi \lambda} N_u(H_2) A_{ul} ~~~~
ergs\,s^{-1}\,cm^{-2}\,sr^{-1}~,
\end{eqnarray}

{\noindent}where $I_{ul}$ is the integrated intensity of the line,
$\lambda$ is the wavelength of the transition $J=u-l$, $A_{ul}$ is the
spontaneous transition probability, $N_u(H_2)$ is the column density
of the upper rotational level of the transition.

\subsection{Detection of the S(1) line at 17.035 $\mu$m}

We recall here the results of our analysis of the S(1) line. For more
details, we refer the reader to the Paper~I. As shown in
Fig.~\ref{HD97} (top panel), we have detected the H$_2$ pure
rotational S(1) line near 17.03 $\mu$m. In the flux-calibrated
spectrum, the standard deviation ($\sigma$) of the continuum flux was
calculated in regions less influenced by telluric absorption, and
close to the feature of interest. We deduced a 6$\sigma$ detection in
amplitude for the line, corresponding to a signal-to-noise ratio of
about 11$\sigma$ for the line, when integrating the signal over a
resolution element (6 pixels). The line is not spectrally resolved as
we can fit it with a Gaussian with a full width at half maximum (FWHM)
equal to a spectral resolution element of 21 \kms\
(Fig.~\ref{HD97}). From our fit, we derived the integrated flux in the
line (see Table~\ref{tab_flux}). Once the spectrum is corrected from
the Earth's rotation, and knowing the heliocentric radial velocity of
HD~97048 \citep[+21 \kms;][]{Acke05}, we estimated, from the
wavelength position of the Gaussian peak, the radial velocity of H$_2$
to be about 4$\pm$2 \kms\ in the star's rest frame.  We thus
considered that the radial velocity of the H$_2$ is compatible with
zero (at the \visir\ resolution) and therefore similar to that of the
star, implying that the emitting gas is gravitationally bound to the
star, and likely arising from the disk and not from an outflow. H$_2$
outflows from T Tauri stars usually have (blueshifted) velocities of a
few tens up to a hundred of \kms\ when a clear outflow is identified.
Otherwise, when the line is detected at zero systemic velocity and is
narrow or unresolved as is the case here, it is generally attributed
to the disk \citep[e.g.][] {Herczeg06, Bitner08}. The H$_2$ line is
not resolved spatially either. Given the \visir\ spatial resolution of
about 0.427'' at 17.03 $\mu$m, and the star distance (180 pc from the
Sun), we can assess that the emitting H$_2$ is located within the
inner 35 AU of the disk.  Assuming that the H$_2$ gas follows the same
(Keplerian) kinematics as the disk, the emitting gas observed with
\visir\ is likely not concentrated significantly in the innermost AU
of the disk ($<$ 5~AU) otherwise rotational broadening would be
observed.  Indeed, near the central star, the rotational velocity of
the disk is of the order of a hundred of \kms, to be compared to the
spectral resolution of VISIR, 21 \kms. The emitting H$_2$ is thus more
likely distributed in an extended region within the inner disk,
between 5~AU and 35~AU of the disk. Assuming the emission arises from
an isothermal mass of optically thin H$_2$, we estimated the
corresponding column densities and masses of H$_2$ as a function of
prescribed temperatures (for 150~K, 300~K and 1000~K; see Paper~I for
details about the method). In Table~\ref{tab_flux} are reported the
integrated flux of the S(1) line, its intensity, and the corresponding
column density for the $J=3$ rotational level. Since we have a
signal-to-noise ratio of about 11$\sigma$ for the line, we can deduce
9\% error bars on theses values (see Table~\ref{tab_flux}).

\subsection{Physical properties of H$_2$}

\begin{figure*}[!htbp]
\begin{center}
\includegraphics[height=7cm]{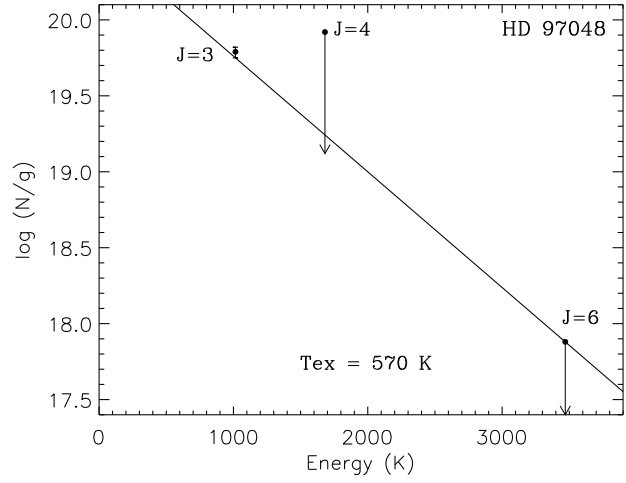}
\caption{\footnotesize Excitation diagram for H$_2$ towards
  HD~97048. If the three rotational levels are populated by thermal
  collisions, their populations follow the Boltzmann law, and the
  upper limit on the excitation temperature is given by considering
  the lower limit on the column density of the $J=3$ level and the
  upper limit on the $J=6$ population level (solid line). Thus, in
  this case, the gas temperature should be lower than 570~K.}
\label{diag_ex}
\end{center}
\end{figure*}

The estimates of the column densities of the $J=3$, $J=4$, and $J=6$
rotational levels of H$_2$ allowed us to plot the excitation diagram
of H$_2$ (see Fig.~\ref{diag_ex}). Assuming that all three levels are
populated by thermal collisions (LTE), we estimated the excitation
temperature of the observed gas. The upper limit on the column density
of the $J=6$ level strongly constrains the value of the column density
of the $J=4$ level. Indeed, the population of the $J=4$ level should
correspond to the excitation temperature given by the ratio of the
column densities of the $J=3$ and $J=6$ levels. We assumed that the
population of the $J=3$ and $J=6$ levels follow the Boltzmann law,
which corresponds to a linear fit to the points on the excitation
diagram (Fig.~\ref{diag_ex}):

\begin{eqnarray}
\frac{N(H_2)_{J=3}}{N(H_2)_{J=6}} = \frac{g_3}{g_6} \times exp \Big(-\frac{E_3 - E_6}{k ~T_{ex}}~ \Big), 
\end{eqnarray}

{\noindent}where $N(H_2)_{J=i}$ is the column density, $g_i$ is the
statistical weight, and $E_i$ is the energy of the $J=i$ level; $k$ is
the Boltzmann constant and $T_{ex}$ is the temperature defined as the
excitation temperature. Since we only have upper limits on the column
densities of the $J=4$ and $J=6$ levels, only an upper limit on the
excitation temperature is relevant. Since the temperature is inversely
proportional to the slope on the excitation diagram, in order to
obtain the upper limit on the excitation temperature, we considered
the lower value of the column density of the $J=3$ level (i.e., the
measured value minus 1$\sigma$) and the upper limit on the $J=6$
population level (that procedure yields the minimum slope / maximum
temperature and corresponds to the solid line on Fig.~\ref{diag_ex}).
We thus find that the excitation temperature of H$_2$ should be lower
than 570~K. In this case, the column density of the $J=4$ level should
be lower than 1.6$\times$10$^{20}$ cm$^{-2}$ which is lower than the
value obtained from the 3$\sigma$ upper limit (see
Table~\ref{tab_flux}).

Under the assumption that the H$_2$ emission is optically thin, that
the emitting H$_2$ is in LTE at a temperature of about 570~K, and that
the source size is equal or smaller than \visir 's beam size, we
derived upper limits on the H$_2$ mass (see Table~\ref{tab_flux}):

\begin{eqnarray}
M_{gas}=f\times 1.76\times 10^{-20}\frac{F_{ul} d^2}{(hc / 4 \pi
  \lambda) A_{ul} x_u(T)}~~M_{\odot}~, 
\end{eqnarray}

{\noindent}where $F_{ul}$ is the line flux, $d$ the distance in pc to
the star, $x_u(T)$ is the fractional population of the level $u$ at
the temperature $T$ in LTE \citep[for details on the calculation
method, see][]{Van_Dishoeck_92}, $f$ is the conversion factor required
for deriving the total H$_2$ gas mass from the H$_2$-ortho or
H$_2$-para mass. Since M$_{H2}$=M$_{H2}$(ortho)+M$_{H2}$(para), then
$f$ =1+1/(ortho/para) for the S(1) line (a H$_2$-ortho transition) and
$f$ =1+ortho/para for the S(2) and S(4) lines (H$_2$-para
transitions).  The equilibrium ortho-para ratio at the temperature T
was computed using Eq.(1) from \citet{Takahashi01b}.

%-------------------------------------------------------------------------------
\section{Discussion}
\label{concl}

Our previous high-resolution spectroscopic observations with \visir\
of the S(1) pure rotational line of H$_2$ at 17.035 $\mu$m of HD~97048
revealed the presence of significant amounts of warm gas in the
surface layer of the disk in the inner 35 AU from the central star
(Paper~I). This detection confirmed that HD~97048 is a young object
surrounded by a circumstellar disk at an early stage of evolution,
still rich in gas.

In the present paper, we report the non-detections of the S(2) and
S(4) lines of H$_2$ in the disk of the Herbig Ae star HD~97048. Using
\visir\ high-resolution spectra, we compute upper limits for the
column densities of the $J=4$ and $J=6$ rotational levels. Using the
results obtained from the analysis of the S(1) line (column density of
the $J=3$ level), and assuming that the $J=3$ to $J=6$ levels are
excited by thermal collisions we derive an upper limit on the
excitation temperature of H$_2$ of about 570~K.

This constraint on a temperature lower than 570~K for the gas allows
us to set a stronger constraint on the mass of warm H$_2$ ($<$ 0.1
M$_{\rm Jup}$; 1 M$_{\rm Jup}$ $\sim$ 10$^{-3}$ M$_{\odot}$) in the
inner 35 AU of the disk than that estimated in Paper~I. Indeed, in
Paper~I, we derived masses of the warm gas in the range from 10$^{-2}$
to nearly 1 M$_{\rm Jup}$, for temperatures in the range from 150~K to
1000~K and assuming LTE.  The mass of warm gas we estimate is also
lower than that we derived from the paper by \cite{Lagage06} who
estimated the mass of H$_2$ in the disk to be of about 0.01
M$_{\odot}$. By assuming that the surface density $\Sigma$ follows a
power law $\Sigma (r) = \Sigma (r/370)^q$ with an index q equal to
$-3/2$ \citep{Lagage06}, we estimated in Paper~I a minimum mass of gas
in the inner 35 AU of the disk to be of the order of 3 M$_{\rm
  Jup}$. But it should be pointed out that mid-IR H$_2$ lines are only
probing warm gas located in the surface layer of the disk, when a
higher mass of colder gas is expected to be present in the interior
layers of the disk.

  However, from their disk model, \citet{Carmona08} concluded that the
  expected peak flux of the S(1) line at 17.035 $\mu$m, observed at a
  spectral resolution of 20\,000, should be less than 0.3\%\ of that
  of the continuum at temperatures higher than 150 K, and thus should
  not be observable with the existing instruments. Those authors used
  a two-layer model \citep{Chiang97, DULLEMOND01} of a gas-rich disk
  (column density of N(H$_2$)=10$^{23}$ cm$^{-2}$) seen face-on,
  located at 140 pc from the Sun, with LTE for the gas and dust,
  T$_{gas}$=T$_{dust}$, and assumed a constant gas-to-dust mass ratio
  about 100.

  As shown in Paper~I from the observation of the S(1) line in the
  disk of HD~97048, such a detection can only be explained if the
  physical conditions of the gas differ from those used by
  \citet{Carmona08}. In Paper~I, we proposed several ways to explain
  our detection. First, assuming equal dust and gas temperatures
  (LTE), we estimated gas-to-dust mass ratios much larger than the
  canonical value of 100.  This hypothesis can not be ruled out by the
  non-detections of the S(2) and S(4) lines. One possible
  interpretation to explain a very high gas-to-dust ratio is that the
  dust is partially depleted from the uppermost disk surface layer,
  where the H$_2$ emission originates. The spatial decoupling between
  the gas and the dust may be due to low densities in the surface
  layers, or dust settling and coagulation into larger particles. The
  physical conditions may thus rapidly differ from the LTE ones,
  i.e. $T_{gas} > T_{dust}$. In the upper disk surface layers,
  photoelectric heating can thus play a significant role in the gas
  heating process \citep{Kamp04, Jonkheid07}. In addition,
  photoelectric heating will be much efficient on small grains, such
  as PAHs, which show significant features in the different
  observations of HD~97048 \citep[e.g.][and reference
  therein]{Doucet07}. Our upper limit on the excitation temperature is
  consistent with photoelectric heating. Indeed, as shown by
  \cite{Kamp04}, photoelectric heating is characterized by a
  temperature limit ($\sim$1000~K) since for high stellar UV flux,
  grain ionization can be so large that the photoelectric effect is
  less efficient and $T_{gas}$ does not increase anymore.

However, our observations do not allow us to rule out the possibility
of a gas heating due to other excitation mechanisms such as UV pumping
or X-ray heating. Indeed, UV and X-ray heating are likely
possibilities that can heat the gas to temperatures signiﬁcantly
hotter than the dust. \cite{Nomura05} modeled UV heating of
circumstellar disks around T Tauri stars and the resulting H$_2$
emission. In their model the temperature can be of about 1000~K at 10
AU from the central star, and the predicted H$_2$ S(2) line flux is
about 2$\times$10$^{-15}$ \flux. This value is lower than our upper
limit on the S(2) line flux, but their model was for a less massive
star and disk.

The upper limit on the excitation temperature of about 570~K we found
for H$_2$ in the disk of HD~97048, could also be consistent with X-ray
heating. Very recently, \cite{Ercolano08} demonstrated from their
two-dimensional photoionization and dust radiative transfer models of
a disk irradiated by X-rays from a T Tauri star, that the uppermost
layers of gas in the disk could reach temperatures of 10$^6$~K at
small radii ($<$0.1 AU) and 10$^4$~K at a distance of 1 AU. The gas
temperatures decrease sharply with depth, but appear to be completely
decoupled from dust temperatures down to a column depth of
$\sim$5$\times$10$^{21}$ cm$^{-2}$. These results are consistent with
those of \cite{Glassgold07} who computed that at 20 AU from the
central star, the temperature can reach 3000~K at the surface of the
disk before dropping to 500-2000~K in a transition zone and then to
much cooler temperatures deep in the disk. By extrapolation, these
models suggest that in the inner 35 AU from the central star, the
upper layers of the disk could reach temperatures consistent with the
upper limit of about 570~K we find for HD~97048.

The kinetic temperature of the gas is generally supposed to be given
by the population of the first rotational levels (namely $J=0$ and
$J=1$) since their critical densities are relatively low, and can not
be higher than the excitation temperature given by the higher energy
levels. Thus, in any case, even if the H$_2$ we observe around
HD~97048 is not essentially excited by thermal collisions, the
populations of the three levels derived from our observations give
strong constraints on the gas kinetic temperature.  Our upper limit on
the temperature of about 570~K is thus reliable whatever the
mechanisms responsible for the excitation of the observed gas. We
stress that the excitation temperature we find for H$_2$ from the
\visir\ observations of HD~97048 is close to that derived from the
TEXES observations of AB Aur (670~K) by \cite{Bitner07}, which
reinforces the similarities between the two stars (see Paper~I).

In order to better constrain the physical conditions of H$_2$,
high-sensitivity and high-resolution space-based spectrographs in the
mid-IR would be required \citep[e.g.][]{Boulanger08}. In particular,
only the observation of the S(0) emission line near 28$\mu$m would
allow us to clearly constrain both the kinetic temperature of the
observed gas and the physical mechanisms responsible its excitation.

%%%%%%%%%%%%%%%%%%%%%%%%%%%%%%%%%%%%%%%%%%%%%%%%%%%%%%%%%%%%%%%%%%%%%%%%%%
\acknowledgments This work is based on observations obtained at
ESO/VLT (Paranal) with \visir, programs' number 079.C-0839A and
079.C-0839B. CMZ is supported by a CNES fellowship.  This work was
supported by \emph{Agence Nationale de la Recherche} (ANR) of France
through contract ANR-07-BLAN-0221. We also thank \emph{Programme
  National de Physique Stellaire} (PNPS) of CNRS/INSU, France, for
supporting part of this research.


%% in parentheses, after the keyword, but they will not be verified.

{\it Facilities:} \facility{ESO/VLT VISIR}

%%%%%%%%%%%%%%%%%%%%%%%%%%%%%%%%%%%%%%%%%%%%%%%%%%%%%%%%%%%%%%%%%%%%%%%%%%%%

\clearpage
%%%%%%%%%%%%%%%%%%%%%%%%%%%%%%%%%%%%%%%%%%%%%%%%%%%%%%%%%%%%%%%%%%%%%%%%%%%%

\newpage

\clearpage

\begin{table*}
\begin{tiny}
\begin{center}
  \caption{Summary of the observations. The airmass and seeing
    intervals are given from the beginning to the end of the
    observations. }
\begin{tabular}{ccccccccccccccccccc}
  \hline
  \hline    
  $\lambda$ & $t_{exp}$ & Airmass    & Optical & Slit     & $R$       & Stand.   & Airmass    & Optical  & Asteroid & Airmass    & Optical \\
  ($\mu$m)    & (s)      &            & seeing  & ('')     &           & Star      &            & seeing    &         &            & seeing \\
  &          &            &  ('')   &          &           &           &            & ('')      &         &            & ('')    \\
  \noalign{\medskip}\hline
  \hline    
  17.0348  & 1800     & 1.72-1.79  & 0.52-0.66  &  0.75   & 14\,000  & HD89388  & 1.90-1.94 & 0.60-0.72 & CERES   & 1.68-1.82 & 0.58-0.69  \\
  12.2786  &  960     & 1.81-1.87  & 1.97-2.16  &  0.4    & 13\,600  & HD91056  & 1.82-1.87 & 1.39-1.82 & VESTA   & 1.03-1.04 & 1.67-2.23 \\
  8.0250   & 1872     & 1.66-1.69  & 0.51-0.65  &  0.4    & 13\,300  & HD92305  & 1.69-1.70 & 0.64-0.65 & VESTA   & 1.54-1.59 & 0.75-0.79 \\
  \hline
\end{tabular}
\label{tab_obs}
\end{center}
\end{tiny}
\end{table*}

\begin{table*}[!htbp]
%\begin{sidewaystable*}[!htbp]
\begin{center}
\begin{tiny}
  \caption{Fluxes, intensities, luminosities and column densities of
    each observed mid-IR transitions of H$_2$. The mass of H$_2$ is
    calculated for a temperature of about 570~K. $\lambda _{obs}$ is
    the wavelength of the centroid of the line. $v_{up}$ and $J_{up}$
    are respectively the vibrational and rotational upper levels of
    the transition of interest.}
\begin{tabular}{ccccccccccccccccccc}
  \noalign{\medskip}\hline
  \hline
  $\lambda _{obs}$ & Transition & $v_{up}$ & $J_{up}$ & Continuum &  Integrated                 & Line                                 & Line              & $N_{J_{up}}({\rm H}_2)$ & M(H$_2$)  \\
                  &            &          &         & flux$^a$   & line flux                   & intensity                            & luminosity        &                       &        \\
  ($\mu$m)        &            &          &         &   (Jy)    &    (\flux)                  & (ergs\,cm$^{-2}$\,s$^{-1}$\,sr$^{-1}$) & log($L/L_{\odot}$) & (cm$^{-2}$)           & (M$_{\rm Jup}$)      \\
  \noalign{\medskip}\hline
  \hline    
  17.03625        &  S(1)      &  0       &  3      &  11.95$\pm$0.94 & 2.4$\pm$0.2$\times$10$^{-14}$ & 5.7$\pm$0.5$\times$10$^{-3}$     & -4.61$\pm 0.04$  & 1.29$\pm$0.12$\times$10$^{21}$ & 1.6$\pm$0.1$\times$10$^{-2}$   \\
  12.2798         &  S(2)      &  0       &  4      &  4.3$\pm$3.0 & $<$5.5$\times$10$^{-14}$      & $<$2.6$\times$10$^{-2}$              & $<$-4.25         & $<$7.46$\times$10$^{20}$ & $<$1.0$\times$10$^{-1}$  \\
  8.0254          &  S(4)      &  0       &  6      &  9.5$\pm$1.1 & $<$4.7$\times$10$^{-15}$      & $<$5.1$\times$10$^{-3}$              & $<$-5.32         & $<$9.80$\times$10$^{18}$ & $<$1.0$\times$10$^{-2}$  \\
  \hline
\end{tabular}
\label{tab_flux}
\begin{list}{}{} 
  $^a$: The error bars on the continuum fluxes are 3$\sigma$ error
  bars.
\end{list}
\end{tiny}
\end{center}
%\end{sidewaystable*}
\end{table*}

\end{document}